\documentclass[apj]{emulateapj}
 
\newcommand{\kms}       {\mbox{km s$^{-1}$}}%
\newcommand{\kmsMpc}    {\mbox{km s$^{-1}$ Mpc$^{-1}$}}%

\newcommand{\msun}      {\mbox{$M_\odot$}} 

\shortauthors{Kannappan \& Gawiser}
\shorttitle{Galaxy Stellar Mass Uncertainties}

\begin{document}
 
\title{Systematic Uncertainties in Stellar Mass Estimation for Distinct Galaxy Populations}
\author{Sheila J. Kannappan\altaffilmark{1,2} and
	Eric Gawiser\altaffilmark{1,3}}
\altaffiltext{1}{NSF Astronomy \& Astrophysics Postdoctoral Fellow}
\altaffiltext{2}{Department of Astronomy, The University of Texas at
        Austin, 1 University Station C1400, Austin, TX 78712-0259;
        sheila@astro.as.utexas.edu}
\altaffiltext{3}{Department of Astronomy, Yale University, PO Box 208101, 
  New Haven, CT 06520; gawiser@astro.yale.edu}

\begin{abstract}
We show that different stellar-mass estimation methods yield overall
mass scales that disagree by factors up to $\sim$2 for the $z=0$
galaxy population, and more importantly, {\it relative} mass scales
that sometimes disagree by factors $\ga$3 between distinct classes of
galaxies (spiral/irregular types, classical E/S0s, and E/S0s whose
colors reflect recent star formation).  This comparison considers
stellar mass estimates based on (a) two different calibrations of the
correlation between $K$-band mass-to-light ratio and $B-R$ color (Bell
et al., Portinari et al.) and (b) detailed fitting of {\it UBRJHK}
photometry and optical spectrophotometry using two different population
synthesis models (Bruzual-Charlot, Maraston), with the same initial
mass function in all cases.  We also compare stellar+gas
masses with dynamical masses.  This analysis offers only weak
arguments for preferring a particular stellar-mass estimation method,
given the plausibility of real variations in dynamical properties and
dark matter content.  These results help to calibrate the systematic
uncertainties inherent in mass-based evolutionary studies of galaxies,
including comparisons of low and high redshift galaxies.
\end{abstract}

\keywords{galaxies: evolution}

\section{\sc Introduction}

The modern trend toward studying galaxy properties as a function of
mass rather than luminosity has led to remarkable advances in our
understanding of galaxy evolution, making the calibration of mass
estimation techniques a high priority.  Based on comparisons of
dynamical and stellar-mass ($M_{dyn}$ and $M_*$) estimates,
\citet{drory.bender.ea:comparing} and
\citet{rettura.rosati.ea:comparing} argue that multi-band photometry
alone can provide accurate $M_*$ estimates that correlate well with
$M_{dyn}$.  Even better, modeling of the correlation between optical
colors and stellar mass-to-light ratios $M_*/L$ suggests that
factor-of-two accuracy in $M_*$ may be achievable with just three
filters, especially when using optical colors to infer an $I$- or
$K$-band $M_*/L$ \citep{bell.:stellar,portinari.sommer-larsen.ea:on}.
However, recent work \citep[][]{maraston:evolutionary} cautions that
$M_*$ estimation may be more complicated than previously assumed, as
young stellar populations may contribute substantially to not only
optical but also near-infrared light, via thermally pulsing asymptotic
giant branch (TP-AGB) stars.  Using these models and those of
\citet{bruzual.charlot:stellar}, \citet{.franx.ea:comparing} find
substantial inconsistencies between $M_{dyn}$ and $M_*$ estimates at low
and high $z$ that appear only when near-IR photometry is used.

To date, empirical examinations of these issues have relied on mixed
data sets, making it hard to isolate systematics in $M_*$ estimation
from evolution between low and high redshift galaxies and/or effects
of inhomogeneous data.  Here, we take advantage of the high-quality,
uniform data available for the Nearby Field Galaxy Survey
\citep[NFGS,][]{jansen.franx.ea:surface}, including photometry,
spectrophotometry, and gas and stellar kinematics, to evaluate $M_*$
estimation techniques.  Our sample allows us to explore effects of
stellar population age on $M_*$ estimation at a single redshift, as it
includes late-type galaxies, classical E/S0 galaxies that fall on the
red color-$M_*$ sequence, and galaxies with E/S0 morphologies that
fall on the blue color-$M_*$ sequence due to recent star formation
\citep[``blue-sequence E/S0s''][ hereafter
KGB]{kannappan.guie.ea:es0}.

\section{\sc Methods}
\label{sec:methods}

The NFGS provides a broadly representative galaxy sample spanning a
wide range of luminosities and morphologies.  For $M_*$ estimation, we
analyze 141 NFGS galaxies with {\it UBRJHK} photometry and integrated
(slit-scanned) optical spectrophotometry from
\citet{jansen.franx.ea:surface,jansen.fabricant.ea:spectrophotometry}
and the Two Micron All-Sky Survey Extended Source Catalog \citep[2MASS
XSC,][]{jarrett.chester.ea:2mass}; see KGB for sample selection
details. We also check our results using $ugr$ photometry for 92 of
these galaxies, taken from the Sloan Digital Sky Survey
\citep[SDSS,][]{adelman-mccarthy.agueros.ea:fourth}.  Photometry and
spectra are corrected for foreground extinction using
\citet{schlegel.finkbeiner.ea:maps} and the Galactic extinction curve
of \citet{odonnell:rnu-dependent}.  We do not apply internal
extinction corrections to the data used for mass determination (though
corrections based on the method of \citealt{tully.pierce.ea:global}
are used incidentally for defining the red and blue sequences, with no
effect on our mass error budget; see KGB).  However, dust is either
included in our modeling or, in the case of color-$M_*/L$ relations,
neglected following standard practice. We add 0.1 mag in quadrature to
the catalogued internal magnitude uncertainties for all passbands to
account for systematic uncertainties in foreground extinction
corrections, automated 2MASS photometry \citep[see][ hereafter
B03]{bell.mcintosh.ea:optical}, and relative photometric zero points.
For the spectra, we add relative flux calibration uncertainties
(typically 6\%, but up to 9\% outside 4000--6800 \AA; R. Jansen,
priv. comm.) in quadrature to the formal uncertainties.

Our ``reference'' $M_*$ values are computed by fitting the photometry
and spectra to a discrete grid of stellar population synthesis models
from \citet{bruzual.charlot:stellar}, scaled to a ``diet Salpeter''
IMF as used by B03.  We combine two simple stellar populations (SSPs)
in varying mass fractions (100:0\%, 90:10\%, 80:20\%, etc.), where we
have normalized each SSP to $M_*$ = 1 \msun.  Each individual SSP has
one of eight ages (0.025, 0.1, 0.29, 0.64, 1, 2.5, 5, 11 Gyr) and
three metallicities (0.4Z$_\odot$, Z$_\odot$, 2.5Z$_\odot$), and each
combination of SSPs has one of eleven dust optical depths ($\tau_{\rm
V, gas}$ = 0, 0.12, 0.24,..., 1.2).  We derive model photometry by
convolving NFGS (standard Johnson-Cousins), SDSS, and 2MASS filter
profiles with Bruzual-Charlot model spectra and adding attenuation
using a \citet{calzetti:dust} law.  The code scales each model to the
observed photometry in $L_\odot$ units, yielding the estimated $M_*$
for that model, then computes likelihoods $\propto
e^{-\chi^2_{overall}/2}$ for the entire grid of models.  In a first
pass, we fit only the photometry, redshifting the models to match the
individual galaxy spectroscopic redshifts and comparing with
redshift-zero models to determine $k$-corrections.  In a second pass,
the likelihood-weighted average $k$-corrections are applied to the
input photometry, and the code fits both the photometry and the
de-redshifted spectra to a fixed set of models in the rest frame.  We
mask emission lines, limit the spectral range to 3800--7000 \AA, and
convolve the model spectra to the 6\AA\ resolution of the NFGS
spectra.  As the spectra lack absolute flux calibration, their scale
factors are allowed to vary freely.  The likelihood of each model is
the product of the likelihoods inferred from the photometry and the
spectra, so the $\chi^2$ terms sum in the exponent and can be weighted
to set the relative influence of the spectra and photometry.  We adopt
$\chi^2_{overall}=max(\chi^2_{phot},n_{dof})+\chi^2_{spec-raw}/1000 +
\chi^2_{spec-norm}/1000$, where the latter two terms are contributions
from fits to the raw and continuum-normalized spectra and $n_{dof}$ is
the number of degrees of freedom in $\chi^2_{phot}$, normally five
when fitting six filters (losing one to the scale-factor
determination).  Once the photometric data are reasonably well fit
($\chi^2_{phot}\leq n_{dof}$), the likelihoods are affected only by
the spectra.  Otherwise, the likelihoods are equally influenced by the
reduced-$\chi^2$ values of the spectra and photometry, because the
ratio of the number of data points is $\sim$500 (where the strong
covariance between $\chi^2_{spec-raw}$ and $\chi^2_{spec-norm}$
justifies treating them as a joint $\chi^2_{spec}$ term).  Following
\citet{bundy.ellis.ea:mass*1}, we adopt the median of the likelihood
distribution binned over $\log{M_*}$ rather than the best fit to
determine the final $M_*$, and we estimate uncertainties from the 68\%
confidence interval in $\log{M_*}$ (binning in 0.02 dex intervals).

$M_*$ estimates based on the $B-R$ vs.\ $M_*/L_K$ relation are derived from
the calibrations of B03 and \citet[][ hereafter
P04]{portinari.sommer-larsen.ea:on}.  The B03 calibration is based on a
global linear fit to $M_*/L_K$ vs.\ synthetic $B-R$ for a large sample of
galaxies with $ugrizK$ data, where each galaxy is fitted with PEGASE
population synthesis models \citep{fioc.rocca-volmerange:pegase} to find
the best-fit metallicity and exponential star formation history (SFH),
which may be decaying, constant, or rising.  The P04 calibration is
predicted from chemo-photometric models of galactic disks, which include
TP-AGB stars.  We use P04's Salpeter IMF calibration, multiplying the
resulting masses by 0.7 to match the diet Salpeter IMF scale of B03.
Technically, P04 limit their calibration to $B-R$ = 0.95--1.45, and our use
of the relation sometimes extends outside this range.  Factor of two
uncertainties are predicted for color-based mass estimation, primarily due
to variations in SFH but also due to the neglect of internal reddening and
extinction \citep[expected to vary mainly along the $B-R$ vs.\ $M_*/L_K$
relation, see][]{bell.:stellar}.  Because color-$M_*/L$ relations do not
provide a self-consistent way to compute $k$-corrections, we $k$-correct
the input magnitudes using our standard method described above.

When comparing $M_*$ and $M_{dyn}$, we apply more restrictive sample
selection criteria.  After rejecting galaxies flagged as
morphologically peculiar by \citet[][ hereafter
KFF]{kannappan.fabricant.ea:physical}, we define two subsamples for
which mass estimates should be robust: (1) 38 spiral/irregular
galaxies with both HI data from the HyperLeda homogenized {H\,{\sc i}}
catalog \citep{paturel.theureau.ea:hyperleda} and optical
emission-line rotation curves passing the quality criteria of KFF,
with the latter also having asymmetry $<$10\% and extent
$>$1.3$\times$ the $B$-band half-light radius $r_e^B$
\citep[][]{kannappan.barton:tools}; and (2) 26 E/S0 galaxies with
optical (Mg-triplet region) stellar velocity dispersions in the NFGS
database
\citep{kannappan.fabricant:broad,kannappan.fabricant.ea:kinematics},
initially measured within $r_e^B$/4 using the Fourier-space fitting
code of \citet{.franx:new} and rescaled to the $R$-band half-light
radius $r_e^R$ using eqn.~1 of
\citet[][]{cappellari.bacon.ea:sauron}. We require the rescaled
dispersion to satisfy $\sigma_{r_e^R}>120$ \kms\ to ensure negligible
rotation corrections (e.g., Fig.~8 of KGB).

For late types, gas masses are computed from the HI line flux with a
helium-mass correction factor of 1.4 and a type- and mass-dependent
molecular-gas correction factor of 1.06--1.4 \citep[based
on][]{casoli.sauty.ea:molecular}.  We adopt an uncertainty of 50\% in
the total gas mass.  For the E/S0s, which are all massive given our
cut in $\sigma_{r_e^R}$, we assume gas masses are negligible (the
largest measured gas-to-stellar mass ratio in this subsample is
$\la15$\%).

Dynamical masses are computed for the E/S0 subsample using
$M_{dyn}=5r_e^R\sigma_{r_e^R}^2$, as recommended by
\citet{cappellari.bacon.ea:sauron}.  We add 5\% in quadrature to the
uncertainties in $r_e^R$ for profile extrapolation errors, and 5\% and
7\%, respectively, to the uncertainties in $\sigma_{r_e^R}$ for
template mismatch and aperture rescaling errors.  For the
spiral/irregular subsample we use $M_{dyn}=2.5r_e^R V_{rot}^2$,
assuming $V_{rot}$ scales like $\sqrt{2}\sigma$
\citep[e.g.,][]{burstein.bender.ea:global}. Here $V_{rot}$ is
0.5$\times$ the inclination-corrected $W^i_{V_{pmm}}$ linewidth
parameter of KFF.

We assume $H_0=70$ \kmsMpc\ and $d=cz/H_0$.
Note that our error bars do not include distance errors because
distances are largely irrelevant to comparisons between mass
estimates.

\section{\sc Results}
\label{sec:results}

\begin{figure*}[t]
\epsscale{1.1}
\plotone{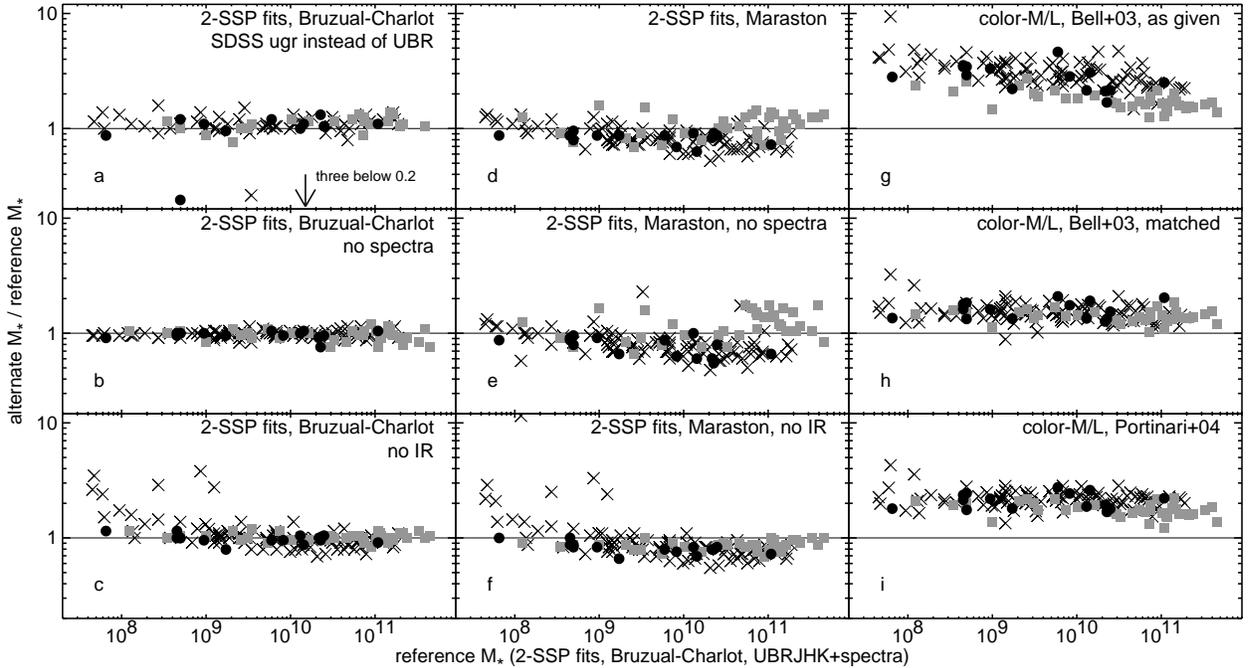}
\caption{Comparison of $M_*$ estimates obtained by various methods,
with symbols coded to show spiral/irregular types (crosses),
red-sequence E/S0s (gray squares), and blue-sequence E/S0s (black
dots).  Each panel shows the ratios between $M_*$'s computed by an
alternate method, as noted, and our reference $M_*$'s computed by
fitting Bruzual-Charlot models to {\it UBRJHK}+spectrophotometry, with
ratios plotted as a function of reference $M_*$.}
\end{figure*}

Various $M_*$ estimation methods are compared in Fig.~1, with the
ratio between the alternate and reference $M_*$'s plotted as a
function of reference $M_*$ in each panel.  Fig.~2 compares
$M_{*(+g)}$ estimates against $M_{dyn}$ in the same format, where the
notation $M_{*(+g)}$ indicates a gas mass correction for late types
(\S~\ref{sec:methods}). We stress that differences between methods do
not imply that one or the other is correct, and differences relative
to $M_{dyn}$ must be interpreted carefully, given the potential for
real variation in dark-matter content or structural properties.

Figs.~1a--c and 2a--c test the robustness of our reference $M_*$'s
against substitution of $ugr$ for $UBR$, omission of the spectra from
the fits, and exclusion of the near-IR {\it JHK} data.  Replacing
$UBR$ with $ugr$, a slight offset appears in Fig.~1a.  NFGS photometry
appears more reliable: Fig.~2b demonstrates that $M_{*(+g)}$ estimates based
on SDSS photometry yield greater scatter relative to $M_{dyn}$ for
late-type galaxies, and inspection of the fits reveals that SDSS data
have a fairly high rate of catastrophic errors relative to 2MASS and
NFGS data, perhaps due to systematic errors in defining galaxy
apertures or profiles.  Note that the apparent improvement in scatter
for E/S0 galaxies in Fig.~2b is probably fortuitous: the open squares
mark galaxies from Fig.~2a that do not have SDSS data, showing that
they tend to be the galaxies with the largest scatter.  $M_*$'s
obtained with and without spectra are closely consistent (Fig.~1b) and
compare similarly with dynamical mass (not shown).  Taking advantage
of this result, we have verified that our reference $M_*$'s are
robust to using a finer resolution in mass ratio between the two SSPs:
100:0\%, 98:2\%, 96:4\%, etc, where for computational efficiency only
the photometry is fitted. Finally, omission of near-IR data (Fig.~1c)
produces generally consistent results, with a few outliers.  The
outliers are all late-type galaxies with fairly low surface
brightness, whose 2MASS magnitudes may be underestimated.
Alternatively, in such bursty systems, $M_*/L$ may be overestimated
without the IR data to anchor the fits. These systems do not show
large shifts from Fig.~2a to 2c because their $M_{*(+g)}$ is
gas-dominated.

Substituting \citet{maraston:evolutionary} models for Bruzual-Charlot
models, stronger differences emerge (Figs.~1d--f and 2d--f).  Fig.~1d
shows a factor of two difference in the relative $M_*$ scales of
high-mass red-sequence E/S0s compared to both blue-sequence E/S0s and
late-type galaxies.  With spectra omitted (Fig.~1e), this difference
grows to a factor of three, whereas with near-IR {\it JHK} data
omitted (Fig.~1f), it disappears entirely, though a small overall
scale difference remains (with Maraston models yielding
$\la$1.3$\times$ lower $M_*$).  Excluding near-IR data, $M_{*(+g)}$
estimates based on Bruzual-Charlot and Maraston models compare nearly
identically to $M_{dyn}$ (Fig.~2c--d).  The differences when near-IR
data are included come almost entirely from the Maraston models and
cause shifts relative to $M_{dyn}$ in Fig.~2e--f.  However, these
shifts are within the uncertainties and might be physical: in a
hierarchical scenario, late-type galaxies may have both more dark
matter and more scatter in dark matter content than early-type
galaxies.  Also, blue-sequence E/S0s may have basic structural
differences from red-sequence E/S0s that lead to overestimated
$M_{dyn}$, if their $\sigma$ and/or $r$ values are elevated due to
incomplete post-merger evolution or disk-building processes (KGB).

\begin{figure}
\epsscale{1.1}
\plotone{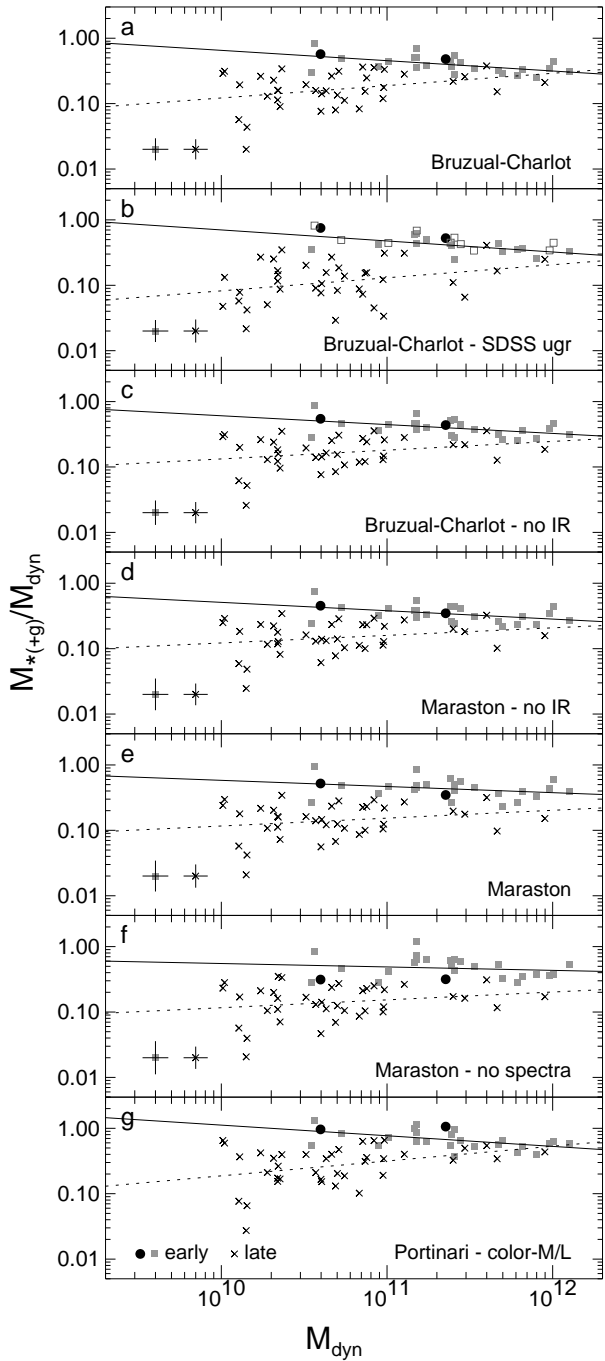}
\caption{Comparison of $M_{*(+g)}$ (gas-mass corrected for late types
only, \S~\ref{sec:methods}) and $M_{dyn}$ for various methods of $M_*$
estimation. Symbols are as in Fig.~1, with the addition of an open
square symbol in panel {\it b} to mark E/S0s from panel {\it a} for
which SDSS $ugr$ data are not available. Least-squares fits and
typical error bars are shown for early and late types, with
blue-sequence E/S0s included with red-sequence E/S0s.}
\end{figure}

Figs.~1g--i and 2g examine $M_*$ estimates based on color-$M_*/L$
relations taken from B03 and P04.  Using the B03 relation as given, we
find strong disagreement compared to our reference $M_*$'s in both
overall mass scale and relative scales between different galaxy
classes, by factors $\ga$3.  However, inspection of the data from
which the B03 relation was determined (their Fig.~20) reveals that
their linear fit is skewed by a two-component distribution, consisting
of a dominant linear locus and a cloud of outliers with blue colors
and high $M_*/L$.  These outliers may be analogous to the outliers we
find in Fig.~1c \& f or may reflect low metallicities (B03).  In any
case, the overall trends can be harmonized if we refit B03's
color-$M_*/L$ relation using only the primary linear locus. Fig.~1h
demonstrates good agreement with only a small overall scale difference
when we adopt the modified relation $\log{M_*/L_K}=-0.616+0.34(B-R)$
for $(B-R)>1.2$ and $\log{M_*/L_K}=-0.808+0.5(B-R)$ for $(B-R)<1.2$,
fitted by eye to their Fig.~20. These formulae include a factor of 1.2
to convert between the $M_*$ scales of the PEGASE and Bruzual-Charlot
models; this offset is noted by B03 when comparing their fit to
earlier predictions from \citet{bell.:stellar} and can be estimated
from their Fig.~20.  In addition to adopting a modified color-$M_*/L$
relation, Fig.~1h adjusts the color-based $M_*$'s for the influence of
dust and starbursts.  B03 estimate that dust and starbursts, if
modeled, would add $\sim$15\% and $\sim$10\% respectively to their
$M_*$ estimates, so we boost $M_*$'s for galaxies with detected
$H\alpha$ by 15\% and $M_*$'s for all galaxies except massive E/S0s
($M_*>10^{11}\msun$) by a separate 10\%.  With this matched-B03
calibration, we find good agreement between the two color-based $M_*$
estimation methods (Fig.~1h--i), with overall scales
$\sim$1.5--1.8$\times$ higher than our reference $M_*$'s, possibly due
to differences in assumed SFHs and/or photometric zero points.  There
may be a slight tendency for the P04 calibration to give higher $M_*$
to younger galaxies, which reduces the offset between E/S0s and
late-type galaxies in comparison to $M_{dyn}$ (Fig.~2g).  This
apparent improvement in the match between $M_{*(+g)}$ and $M_{dyn}$ should
be taken with a grain of salt, as it reflects the loss of information
on why galaxies are blue or red (age, dust, metallicity).


In summary, our results demonstrate systematic uncertainties in $M_*$
estimation corresponding to factors up to $\sim$2 overall and $\ga$3
between distinct galaxy classes, even using our modified B03
calibration and matched IMFs.  Outliers affect B03's original
color-$M_*/L_K$ calibration and also emerge when we compare results
from stellar population modeling with and without IR data (note both
B03 and this work rely on 2MASS).  More generally, $M_*$ estimates are
highly sensitive to IR and spectral information when using Maraston
models, especially for red- and blue-sequence E/S0s.  An Occam's razor
argument might justify preferring Bruzual-Charlot models, which yield
consistent results with or without spectra or IR data. However, we are
unable to find a strong physical argument for preferring a particular
set of models based on comparisons with $M_{dyn}$, and we caution that
agreement between $M_{*(+g)}$ and $M_{dyn}$ is not by itself proof of
better $M_*$ estimation, given evidence for variations in dark matter
content, dynamical state, and age/dust/metallicity.  We conclude that
mass-based evolutionary studies of galaxies should explicitly consider
the potential effects of systematic errors in $M_*$, particularly when
analyzing young and old galaxies together across galaxy classes or
between low and high $z$.

\acknowledgments

We thank A. Baker, D. Christlein, K. Gebhardt, R. Jansen,
I. Labb\'{e}, D. Mar, K. Williams, and M. Wolf for helpful
discussions.  SJK and EG were supported by NSF Astronomy \&
Astrophysics Postdoctoral Fellowships under awards AST-0401547 \&
0201667 respectively.  This research used the HyperLeda {H\,{\sc i}}
catalog (Vizier Catalog VII/238) and data from 2MASS, a joint project
of the U. of Massachusetts and IPAC/Caltech, funded by NASA and the
NSF. It also used data from the Sloan Digital Sky Survey (see full
acknowledgement at http://www.sdss.org/collaboration/credits.html).
\\


\footnotesize
\begin{thebibliography}{28}
\expandafter\ifx\csname natexlab\endcsname\relax\def\natexlab#1{#1}\fi

\bibitem[{{Adelman-McCarthy} {et~al.}(2006){Adelman-McCarthy}, {et~al.}}]{adelman-mccarthy.agueros.ea:fourth}
{Adelman-McCarthy}, J.~K., et al. 2006, \apjs, 162, 38

\bibitem[{{Bell} \& {de Jong}(2001)}]{bell.:stellar}
{Bell}, E.~F. \& {de Jong}, R.~S. 2001, \apj, 550, 212

\bibitem[{{Bell} {et~al.}(2003){Bell}, {McIntosh}, {Katz}, \&
  {Weinberg}}]{bell.mcintosh.ea:optical}
{Bell}, E.~F., {McIntosh}, D.~H., {Katz}, N., \& {Weinberg}, M.~D. 2003, \apjs,
  149, 289 [B03]

\bibitem[{{Bruzual} \& {Charlot}(2003)}]{bruzual.charlot:stellar}
{Bruzual}, G. \& {Charlot}, S. 2003, \mnras, 344, 1000

\bibitem[{{Bundy} {et~al.}(2005){Bundy}, {Ellis}, \&
  {Conselice}}]{bundy.ellis.ea:mass*1}
{Bundy}, K., {Ellis}, R.~S., \& {Conselice}, C.~J. 2005, \apj, 625, 621

\bibitem[{{Burstein} {et~al.}(1997){Burstein}, {Bender}, {Faber}, \&
  {Nolthenius}}]{burstein.bender.ea:global}
{Burstein}, D., {Bender}, R., {Faber}, S., \& {Nolthenius}, R. 1997, \aj, 114,
  1365

\bibitem[{{Calzetti}(2001)}]{calzetti:dust}
{Calzetti}, D. 2001, \pasp, 113, 1449

\bibitem[{{Cappellari} {et~al.}(2006){Cappellari}, {et al.}}]{cappellari.bacon.ea:sauron}
{Cappellari}, M., et al. 2006, \mnras, 366, 1126

\bibitem[{{Casoli} {et~al.}(1998){Casoli}, {et al.}}]{casoli.sauty.ea:molecular}
{Casoli}, F., {Sauty}, S., {Gerin}, M., {Boselli}, A., {Fouque}, P., {Braine},
  J., {Gavazzi}, G., {Lequeux}, J., \& {Dickey}, J. 1998, \aap, 331, 451

\bibitem[{{Drory} {et~al.}(2004){Drory}, {Bender}, \&
  {Hopp}}]{drory.bender.ea:comparing}
{Drory}, N., {Bender}, R., \& {Hopp}, U. 2004, \apjl, 616, L103

\bibitem[{{Fioc} \& {Rocca-Volmerange}(1997)}]{fioc.rocca-volmerange:pegase}
{Fioc}, M. \& {Rocca-Volmerange}, B. 1997, \aap, 326, 950

\bibitem[{{Jansen} {et~al.}(2000{\natexlab{a}}){Jansen}, {Fabricant}, {Franx},
  \& {Caldwell}}]{jansen.fabricant.ea:spectrophotometry}
{Jansen}, R.~A., {Fabricant}, D., {Franx}, M., \& {Caldwell}, N.
  2000{\natexlab{a}}, \apjs, 126, 331

\bibitem[{{Jansen} {et~al.}(2000{\natexlab{b}}){Jansen}, {Franx}, {Fabricant},
  \& {Caldwell}}]{jansen.franx.ea:surface}
{Jansen}, R.~A., {Franx}, M., {Fabricant}, D., \& {Caldwell}, N.
  2000{\natexlab{b}}, \apjs, 126, 271

\bibitem[{{Jarrett} {et~al.}(2000){Jarrett}, {Chester}, {Cutri}, {Schneider},
  {Skrutskie}, \& {Huchra}}]{jarrett.chester.ea:2mass}
{Jarrett}, T.~H., {Chester}, T., {Cutri}, R., {Schneider}, S., {Skrutskie}, M.,
  \& {Huchra}, J.~P. 2000, \aj, 119, 2498

\bibitem[{{Kannappan} \& {Barton}(2004)}]{kannappan.barton:tools}
{Kannappan}, S.~J. \& {Barton}, E.~J. 2004, \aj, 127, 2694

\bibitem[{{Kannappan} \& {Fabricant}(2001)}]{kannappan.fabricant:broad}
{Kannappan}, S.~J. \& {Fabricant}, D.~G. 2001, \aj, 121, 140

\bibitem[{{Kannappan} {et~al.}(2002){Kannappan}, {Fabricant}, \&
  {Franx}}]{kannappan.fabricant.ea:physical}
{Kannappan}, S.~J., {Fabricant}, D.~G., \& {Franx}, M. 2002, \aj, 123, 2358 [KFF]

\bibitem[{{Kannappan} {et~al.}(2006{\natexlab{a}}){Kannappan}, {Guie}, \&
  {Baker}}]{kannappan.guie.ea:es0}
{Kannappan}, S.~J., {Guie}, J.~M., \& {Baker}, A.~J. 2006{\natexlab{a}}, submitted [KGB]

\bibitem[{{Kannappan} {et~al.}(2006{\natexlab{b}}){Kannappan}, {et al.}}]{kannappan.fabricant.ea:kinematics}
{Kannappan}, S.~J., et al. 2006{\natexlab{b}}, in preparation

\bibitem[{{Maraston}(2005)}]{maraston:evolutionary}
{Maraston}, C. 2005, \mnras, 362, 799

\bibitem[{{O'Donnell}(1994)}]{odonnell:rnu-dependent}
{O'Donnell}, J.~E. 1994, \apj, 422, 158

\bibitem[{{Paturel} {et~al.}(2003){Paturel}, {Theureau}, {Bottinelli},
  {Gouguenheim}, {Coudreau-Durand}, {Hallet}, \&
  {Petit}}]{paturel.theureau.ea:hyperleda}
{Paturel}, G., {Theureau}, G., {Bottinelli}, L., {Gouguenheim}, L.,
  {Coudreau-Durand}, N., {Hallet}, N., \& {Petit}, C. 2003, \aap, 412, 57

\bibitem[{{Portinari} {et~al.}(2004){Portinari}, {Sommer-Larsen}, \&
  {Tantalo}}]{portinari.sommer-larsen.ea:on}
{Portinari}, L., {Sommer-Larsen}, J., \& {Tantalo}, R. 2004, \mnras, 347, 691 [P04]

\bibitem[{{Rettura} {et~al.}(2006){Rettura}, {et al.}}]{rettura.rosati.ea:comparing}
{Rettura}, A., et al. 2006, astro-ph/0608545

\bibitem[{{Schlegel} {et~al.}(1998){Schlegel}, {Finkbeiner}, \&
  {Davis}}]{schlegel.finkbeiner.ea:maps}
{Schlegel}, D.~J., {Finkbeiner}, D.~P., \& {Davis}, M. 1998, \apj, 500, 525

\bibitem[{{Tully} {et~al.}(1998){Tully}, {Pierce}, {Huang}, {Saunders},
  {Verheijen}, \& {Witchalls}}]{tully.pierce.ea:global}
{Tully}, R.~B., {Pierce}, M.~J., {Huang}, J., {Saunders}, W., {Verheijen}, M.
  A.~W., \& {Witchalls}, P.~L. 1998, \aj, 115, 2264

\bibitem[{{van der Marel} \& {Franx}(1993)}]{.franx:new}
{van der Marel}, R.~P. \& {Franx}, M. 1993, \apj, 407, 525

\bibitem[{{van der Wel} {et~al.}(2006){van der Wel}, {Franx}, {Wuyts}, {van
  Dokkum}, {Huang}, {Rix}, \& {Illingworth}}]{.franx.ea:comparing}
{van der Wel}, A., {Franx}, M., {Wuyts}, S., {van Dokkum}, P.~G., {Huang}, J.,
  {Rix}, H.-W., \& {Illingworth}, G.~D. 2006, astro-ph/0607649

\end{thebibliography}

\end{document}